\documentclass[]{pasj02} 
\usepackage[switch,mathlines]{lineno} 
\usepackage{natbib} 
\usepackage{rotating}
\usepackage{multirow}
\usepackage{booktabs}
\usepackage[utf8]{inputenc}

\newcommand*{\FirstRevise}[1]{\textcolor{black}{#1}}

\jyear{2024}
\Received{}
\Accepted{}

\begin{document} 

\title{ Origin and Evolution of the $\Omega$ Structure in the Head-Tail Radio Galaxy of Abell~3322 }

\author{
Kohei \textsc{kurahara},\altaffilmark{1,2}\altemailmark\orcid{0000-0003-2955-1239} \email{kurahara.kohei.i7@f.mail.nagoya-u.ac.jp} 
Takuya \textsc{akahori},\altaffilmark{1}\orcid{0000-0001-9399-5331}
Takumi \textsc{ohmura},\altaffilmark{3,4}\orcid{0000-0002-0040-8968}
Shintaro \textsc{yoshiura},\altaffilmark{1}\orcid{0000-0003-0581-5973}
Daisuke \textsc{ito},\altaffilmark{5}\orcid{0009-0000-4742-5098} 
Yik ki \textsc{ma},\altaffilmark{6}\orcid{0000-0003-0742-2006}
Kazuhiro \textsc{nakazawa},\altaffilmark{2,5}\orcid{0000-0003-2930-350X}
Yuki \textsc{omiya},\altaffilmark{5}\orcid{0009-0009-9196-4174}
Kosei \textsc{sakai},\altaffilmark{5}\orcid{????-????-????-????}
Haruka \textsc{sakemi},\altaffilmark{7}\orcid{0000-0002-4037-1346}
and 
Motokazu \textsc{takizawa}\altaffilmark{8}\orcid{0000-0001-9035-7764}
}
\altaffiltext{1}{Mizusawa VLBI Observatory, National Astronomical Observatory of Japan, 2-21-1 Osawa, Mitaka, Tokyo 181-8588, Japan}
\altaffiltext{2}{Kobayashi-Maskawa Institute for the Origin of Particles and the Universe (KMI), Nagoya University, Furo-cho, Chikusa-ku, Nagoya, Aichi 464-8601, Japan}
\altaffiltext{3}{Institute for Cosmic Ray Research, The University of Tokyo, 5-1-5 Kashiwanoha, Kashiwa, Chiba 277-8582, Japan}
\altaffiltext{4}{Division of Science, National Astronomical Observatory of Japan, 2-21-1 Osawa, Mitaka, Tokyo 181-8588, Japan}
\altaffiltext{5}{Departure of Physics, Nagoya University, Furo-cho, Chikusa-ku, Nagoya, Aichi 464-8601, Japan}
\altaffiltext{6}{Max Planck Institute for Radio Astronomy, Bonn, North Rhine-Westphalia, DE, 53121, Germany}
\altaffiltext{7}{Graduate School of Science and Technology for Innovation, Yamaguchi University, 1677-1, Yoshida, Yamaguchi, Yamaguchi 753-0841, Japan}
\altaffiltext{8}{Department of Physics, Yamagata University, Kojirakawa-machi 1-4-12, Yamagata, Yamagata 990-8560, Japan}

\KeyWords{galaxies: clusters: individual (Abell 3322) — radio continuum: galaxies - methods: observational}  

\maketitle

\begin{abstract}
A head-tail galaxy is thought to be a radio galaxy with bent active galactic nuclei (AGN) jets interacting with the intracluster medium (ICM). Study of head-tail galaxies provides us with fruitful insights into the mechanisms of shock waves and turbulence, as well as magnetic-field amplification and cosmic-ray acceleration. A recent MeerKAT observation revealed that a head-tail galaxy in the galaxy cluster, Abell 3322, exhibits a peculiar ``Omega" structure in its shape. In this paper, we investigated this Omega-tail galaxy using the upgraded Giant Meterwave Radio Telescope (GMRT) and the Australia Telescope Compact Array (ATCA). We found that the southern jet tends to be brighter than the northern jet, with a brightness ratio of about 2. This can be attributed to Doppler boost and the inclination of the jets. Our broadband data suggest that the radio spectrum becomes steeper along the jet propagation direction, and the cosmic-ray aging model with a weak reacceleration of cosmic rays is preferable to explain the index profile. We further found a gradient of the spectral index perpendicular to the jet propagation. We discussed the origin of the gradient and suggested that a shock wave along one side of the jets is present. The resultant ram pressure as well as the backflow made at the early stage of the jet may produce the tail component of this Omega-tail galaxy, while the observed Omega-shape structure is more likely due to a twin vortex seen in the low Reynolds number flow.
\end{abstract}


\section{Introduction}\label{sec:1}

A head-tail galaxy is a galaxy possessing an active galactic nucleus (AGN) as a head and (a) bent jet(s) as tail(s). Head-tail galaxies have been categorized into several structural types \citep{1968MNRAS.138....1R}. If the opening angle of the bent jet exceeds 90$^\circ$, it is typically called a wide-angle tail (WAT), while narrower angles \FirstRevise{, less than 90$^\circ$,} are classified as narrow-angle tails (NAT) \citep{2022ApJS..259...31S}. Recent high-sensitivity radio surveys discovered many head-tail galaxies \citep{2011ApJS..194...31P}. For example, 717 new head-tail galaxies have been identified in the VLA FIRST survey at 1.4 GHz \citep{2022ApJS..259...31S}. More recently, the MeerKAT Galaxy Cluster Legacy Survey \citep[MGCLS;][]{2022A&A...657A..56K} observed 115 galaxy clusters in full polarization at L-band (0.9-1.67 GHz) and found a large number of new radio sources, some of which are believed to be head-tail galaxies.

The origin of head-tail galaxies has been investigated in the literature \citep[e.g.,][]{2019A&A...626A...8M, 2020NewAR..8801539H}. Since they are often associated with a galaxy cluster, the tails are thought to be formed by a dynamical or magnetic interaction between the jets and the intra-cluster medium (ICM). A major interaction mechanism is the ram pressure of the ICM  \citep{2020MNRAS.496.4654C, 2023ApJ...951...76O}, indeed some observational studies confirmed that the bending of jets is caused by ram pressure from the merger clusters \citep{1985ApJ...295...80O, 2000MNRAS.311..649S}. The correlation between the morphology of head-tail galaxies and the distance from the cluster center supports this hypothesis \citep[e.g,][]{2025arXiv250517334V}. 

Recent high-sensitivity radio observations have revealed a few rare cases with unique structures. Another proposed scenario for the formation of head-tail galaxy structures involves the tension of magnetic fields in the ICM \citep{2021Natur.593...47C}. They observed a late-stage merging galaxy cluster, Abell 3376, using MeerKAT and found that the bent jets are closely aligned with the cluster's cold front seen in X-rays. They performed MHD simulations and suggested that the bent jets can be formed if there are strong ($\sim 10\mu$G) and coherent ($\sim 100$ kpc) magnetic fields along the cold front. Furthermore, in the galaxy cluster IIZW108, the spatial structure and polarization properties of the bent jets are explained by shear effects and magnetic draping caused by the velocity difference between the AGN jet and the ICM \citep{2022ApJS..259...31S}.

Head-tail galaxy and ICM interaction could generate turbulence and amplify the magnetic field. These turbulence and magnetic fields play an important role in cosmic-ray (CR) (re-)acceleration in galaxy clusters. Recent low-frequency radio observations have reported indirect evidence of re-acceleration phenomena of CR electrons \citep{2022A&A...659A..20I}. In Abell 3376, the spectral aging trend suggests that the jet is colliding with magnetic fields along the cluster's cold front, with electrons being re-accelerated and following the magnetic field lines \citep{2021Natur.593...47C}. This means that the structure and energy properties of head-tail galaxies are very important not only for understanding the origin of their own structure formation but also for studying the properties of the ICM. Head-tail galaxies are good targets for studying particle acceleration, magnetic field amplification, and turbulence generation.

\begin{figure*}[tp]
\begin{center}
\includegraphics[width=16cm]{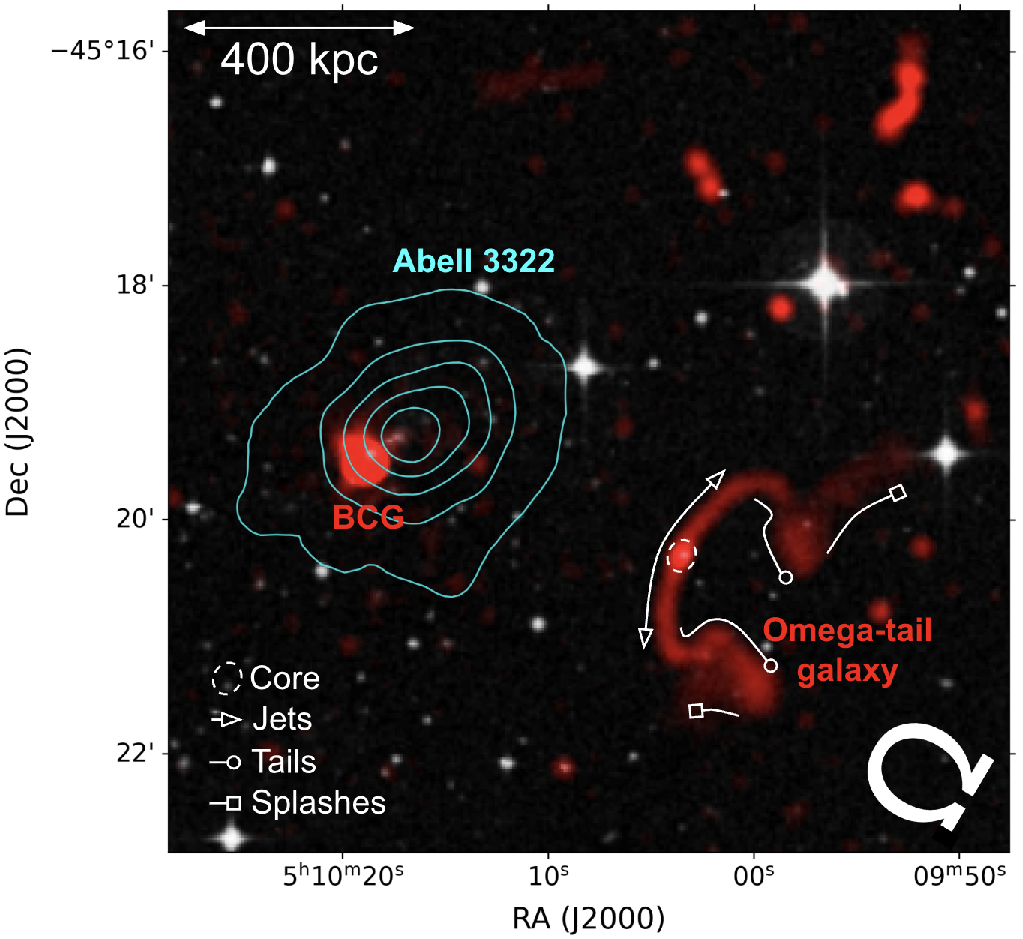}
\end{center}
\caption{A composite image combining X-ray (cyan-contour), optical (white), and radio (red) data. 
The X-ray image is obtained by Chandra data in the 0.5-7.0~keV energy band, smoothed with a Gaussian kernel of $\sigma =$ 23 arcseconds, and the contours are overlaid in cyan and drawn at 6 logarithmically spaced levels between the 90th percentile and the maximum of the smoothed X-ray intensity, after linear normalization. 
The optical data were obtained from the Blue band of the DSS2 survey via SkyView. The radio image corresponds to the frequency-integrated map from the MGCLS survey. The primary target of this study, referred to as the "Omega-tail galaxy", is labeled along with its various structural components. The $\Omega$ symbol is also shown for reference.\\ {Alt text: Composite image of the Omega-tail galaxy showing X-ray (cyan contours), optical (white), and radio (red) data with labeled structures.}}
\label{fig:f00}
\end{figure*}

Abell 3322 is a massive galaxy cluster located at about $z=0.2$. Table \ref{tab:1} summarizes basic parameters of Abell 3322.  Previous X-ray observations \citep{2001A&A...365L...1J} showed that the X-ray surface brightness distribution is elongated with the major axis in the northwest and southeast directions (see also figure \ref{fig:f00}). The central galaxy is spatially offset from the X-ray brightness \FirstRevise{peak position}. This evidence indicates that Abell 3322 is not dynamically well-relaxed yet. Therefore, Abell 3322 is suitable for advancing our understanding in the nature of galaxy clusters under dynamical and thermal relaxation. More recently, this cluster was observed using MeerKAT, and a head-tail galaxy was found at the southwest of the cluster center, though there have been fewer studies detailing the results of the radio observations. Further radio studies of head-tail galaxies \FirstRevise{are} useful for understanding the distribution of matter, both dark and visible, in galaxy clusters, and for elucidating the interaction between AGN jets and the ICM.

In this paper, we report on the results of multi-frequency radio observations of Abell 3322 to detect synchrotron radiation from the head-tail galaxy associated with Abell 3322, including a total of 37 hours of high-resolution and high-sensitivity observations in Bands 3 using uGMRT. We aim to investigate the geometrical structure and the CR electron spectrum of the head-tail galaxy. We discuss the physical properties and roles of the AGN jet interacting with the ICM, by comparing the results with MHD simulations. In Section 2, we describe the details of the observations and the data reduction procedures, and in Section 3, we present some obtained radio images and a spectral index map. In Section 4, we discuss the detailed radio properties of the detected structures and their origins. We have used cosmological parameters $H_0 = 70~{\rm km~s^{-1}~Mpc^{-1}}$, $\Omega_M = 0.3$ and $\Omega_\Lambda = 0.7$ in this work.

\begin{table}[tp]
\tbl{Basic parameters of Abell 3322. }{%
\begin{tabular}{rccc}
\hline
Parameter & Value & Unit & Ref. \\ 
\hline
z & 0.2 & - & (a) \\
$M_{\rm tot}$  & $5.36^{+0.51}_{-0.36}$ & $10^{14}$ M$_{\odot}$ & (a)  \\
$M_{\rm gas}$  & $6.74^{+0.27}_{-0.20}$ & $10^{13}$ M$_{\odot}$ & (a) \\
$kT$ & $5.98^{+0.23}_{-0.23}$ & keV & (a) \\
$kT_{\rm exc}$ & $5.97^{+0.36}_{-0.36}$ & keV & (a) \\
$L_{\rm X}$ 0.1-2.4 keV & $4.91^{+0.18}_{-0.18}$ & $10^{44}$~erg s$^{-1}$ & (a) \\
$L_{\rm X,exc}$ 0.1-2.4 keV & $2.93^{+0.08}_{-0.08}$ & $10^{44}$~erg s$^{-1}$ & (a) \\
$L_{\rm bol}$ 0.01-100 keV & $11.18^{+0.66}_{-0.66}$ & $10^{44}$~erg s$^{-1}$ & (a) \\
$L_{\rm bol,exc}$ 0.01-100 keV & $6.67^{+0.31}_{-0.31}$ & $10^{44}$~erg s$^{-1}$ & (a) \\
$R_{500}$ & 1155 & kpc & (b) \\
$n_e$ & $0.88 \pm 0.01$ & $10^{-2}$ & (b) \\
cusp & $0.19 \pm 0.01$ & - & (b) \\
c & $0.23 \pm 0.01$ & - & (b) \\
w & $1.79 \pm 0.06$ & $10^{-2}$ & (b) \\
{\it Gini} & $0.66 \pm 0.01$ & - & (b) \\
P30 & $0.28 \pm 0.14$ & $10^{-7}$ & (b) \\
P40 & $1.50 \pm 0.73$ & $10^{-8}$ & (b) \\
ell & $ 0.86 \pm 0.01$ & - & (b) \\
Dunamical State & Mix & - & (b) \\ \hline
\end{tabular}}\label{tab:1}
\begin{tabnote}
Refs are (a) \cite{2020ApJ...892..102L}, and (b) \cite{2017ApJ...846...51L}.\\
Row (1): Labels. 
Row (2): Redshift. 
Rows (3)-(4): total and gas mass within R500. 
Rows (5)-(6): core-included and core-excluded cluster temperature. 
Rows (7)-(8): core-included and core-excluded cluster luminosity in the soft band (0.1-2.4 keV). 
Rows (9)-(10): core-included and core-excluded cluster bolometric luminosity (0.01$\sim$100 keV). 
Row (11): ${\rm R}_{500}$; The radius within which the overdensity of the galaxy cluster is 500 times the critical density of the Universe. 
Row (12): Central electron number density. 
Row (13): Cuspiness; Log slope of electron density at radius of $0.04 r_{500}$. 
Row (14): The concentration parameter; It is defined as the ratio of the emission within 2 different circular apertures. 
Row (15): The centroid shift parameter; The variance of the projected separation between the X-ray peak and the centroid of the emission. 
Row (16): The Gini coefficient; If the total flux is equally distributed among the considered pixels, then the Gini coefficient is equal to 0, while if the total flux is concentrated into a single pixel, then its value is equal to 1. 
Rows (17)-(18): Third and fourth order power ratios; A ratio of 2-D multipole decomposition of surface brightness distribution. 
Row (19): Ellipticity. 
Row (20): Classification of galaxy clusters, relaxed (R), mix (M), or disturbed (D). 
\end{tabnote}
\end{table}

\section{Observation and Data Reduction}\label{sec:2}

In this section, we describe our observations, the data used in this paper, and the strategies adopted for our data reduction.

\subsection{uGMRT Observation and Data Reduction}
 
We conducted observations pointing toward the head-tail galaxy of Abell 3322, (RA, DEC) = (05:10:03.1304,$-$45:20:14.719), using the upgraded Giant Metrewave Radio Telescope (uGMRT). We refer to the head-tail galaxy at the pointing center as the Omega-tail galaxy to distinguish it from other head-tail galaxies.

We adopted the wideband mode within Band 3 (300-500 MHz), where the data of the narrowband mode were also recorded simultaneously (project code 42\_044). The center frequency and the bandwidth for the wideband mode were 400 MHz and 200 MHz, respectively, while for the narrowband mode, these values were 317 MHz and 33 MHz, respectively. The observations were performed over three separate days in August and October 2022, utilizing 28 antennas for \FirstRevise{two of the 3 sessions in} October and 26 antennas \FirstRevise{for the only session} in August. Each observation session began with a 10-minute scan of the flux density, bandpass, and polarization calibrators 3C147 and 3C138, 5-minute scans of the phase calibrator J0522-364, and 60-minute scans of the target Abell 3322. At the end of each session, 10-minute scans of the standard calibrators 3C147 and 3C286 were conducted.

The data were processed using Source Peeling and Atmospheric Modeling \citep[SPAM;][]{2014ASInC..13..469I}. The basic data reduction strategy employed with \texttt{SPAM} followed the method described in our previous work \citep{2023PASJ...75S.138K}. In short, we applied direction-dependent calibration \citep[DDC;][]{2017A&A...598A..78I} and used \texttt{WSClean} \citep{2014MNRAS.444..606O} for wide-field, wide-band imaging. \texttt{SPAM} was used to combine the data from all observation days. Pre-calibration was performed on each observation day individually, and subsequent imaging, including {\it source peeling}, was done with the combined UV data from all sessions. For the final imaging, we adjusted the robustness and taper parameters to obtain the best image. The achieved noise level and the dynamic range were optimal with Briggs$ = 0$, so we adopt the Briggs$ = 0$ result in this paper. The \texttt{AIPS} tasks `HGEOM`, `PBCOR`, and `CONVL` were used for data regridding, main beam correction, and beam convolution, respectively.

\subsection{ATCA Observation and Data Reduction}

We conducted observations pointing toward the Omega-tail galaxy using the Australia Telescope Compact Array (ATCA). We adopted the Compact Array Broadband Backend (CABB) correlator, and with the two IFs, we observed at the center frequency of 5000 MHz (C band) with the bandwidth of 2000 MHz and at 9000 MHz (X band) with the bandwidth of 2000 MHz (project code 2022Oct-C3494). We observed the standard calibrator B1934-638 in 10 minutes to obtain flux and bandpass calibration data. Thereafter, we observed the phase calibrator J0514-4554 for 3 minutes and Abell 3322 for 18-minute blocks, alternately. B1934-638 was observed every two hours for flux calibration. The observations were carried out over two days, on January 31 and February 1, 2023. On both days, antenna CA06, which provides the longest baseline length, was out of operation due to maintenance, so the observations were made with five antennas.

The data reduction was performed using MIRIAD software (version 1.5) following the standard ATCA procedures. All data were loaded using the MIRIAD task `ATLOD` with the options `BIRDIE`, `RFIFLAG`, `XYCORR`, and `NOAUTO`. First, the band edges, which are susceptible to bandpass roll-off, were flagged by removing 40 channels using `UVFLAG`. The `UVSPLIT` task was used to divide the data into subsets for the bandpass calibrator, gain/phase calibrator, and the target source. For the bandpass calibrator subset, it was initially trimmed using `PGFLAG` to remove obvious bad data. Further manual flagging of bad data was performed by tasks such as `BLFLAG` and `UVFLAG` for anomalous times and channels while inspecting the data using `UVPLT` and `VARPLT` tasks. Next, the bandpass solution was determined using the `MFCAL` task, and the correction terms were computed by `GPCAL`. These solutions were copied to the gain/phase calibrator subset using `GPCOPY`. The gain/phase calibrator subsets were flagged with the same strategy, and the correction terms were determined by `GPCAL`. Finally, all solutions were applied to the target subset using `GPCOPY` and `GPAVER`, and any remaining bad data were flagged following the same procedures.

Final imaging also used MIRIAD, with tasks INVERT, CLEAN, and RESTOR. `MFCLEAN` was used instead of CLEAN only for Stokes I imaging. A robust parameter of 2 was used to achieve sufficient sensitivity. Task `UVSPLIT` was used for frequency splitting. Again, the AIPS tasks `HGEOM`, `PBCOR`, and `CONVL` were used for data regridding, main beam correction, and beam convolution, respectively.


\begin{table}[htbp]
\tbl{Image parameters.}{%
\begin{tabular}{cccrrc} \hline \hline
Telescope & Frequency & Beam size & Beam PA & R.M.S. & Map$^*$ \\
 & MHz & asec $\times$ asec & degree & $\mu $Jy beam$^{-1}$ & \\  \hline
uGMRT & 400 & $13.2 \times 4.8$ & 177.8 & 25.0 & a \\
ASKAP & 888 & $25.0 \times 25.0$ & 0.0 & 344.6 & b \\
MeerKAT & 1280 & $7.6 \times 7.5$ & 24.2 & 6.2 & c \\
ASKAP & 1368 & $9.7 \times 9.0$ & 1.7 & 148.6 & d \\
ATCA & 5000 & $7.3 \times 3.3$ & 3.9 & 10.8 & e \\
ATCA & 9000 & $4.2 \times 1.8$ & 7.0 & 10.3 & f \\ \hline 
\end{tabular}}\label{t01}
\begin{tabnote}
$^*$ The label in the rightmost column corresponds to the panel in Figure~\ref{f01}.
\end{tabnote}
\end{table}

The parameters of the obtained Stokes I images are shown in Table \ref{t01}. The achieved noise level of the uGMRT data is 25.0 $\mu $Jy~beam$^{-1}$, and the brightest source in the field of view of uGMRT is a galaxy called LEDA~529949, located at 20~arcminutes northwest of the core of the head-tail galaxy, with a peak intensity of 161.0 mJy~beam$^{-1}$, which means the achieved image dynamic range was 6,440. 

For the ATCA data, the achieved noise level is approximately 10 $\mu $Jy~beam$^{-1}$ at both 5 and 9~GHz. The brightest source in the field of view of ATCA is not 2MASX J05101744-4519179 which is described in SIMBAD as the BCG of Abell 3322, but WISEA J051018.77-451927.5 located southeast of the BCG, with peak intensities of 7.3 mJy~beam$^{-1}$ at 5~GHz and 1.4 mJy~beam$^{-1}$ at 9~GHz. The dynamic ranges from these values are approximately 730 and 140, respectively.

\subsection{Other Archival Data}
To examine the radio emission, we checked the Rapid ASKAP Continuum Survey \citep[RACS;][]{2020PASA...37...48M} data, which were downloaded from 
the CSIRO ASKAP Science Data Archive \citep[CASDA;][]{2020ASPC..522..263H}. Although the data showed a similar trend to our high-sensitivity observations, the sensitivity was insufficient, and thus, we did not use them for further discussion.

To investigate the spatial relationship between the host galaxy of the Omega-tail and the surrounding cluster environment, we made use of both optical and X-ray imaging data. The optical data were retrieved from the DSS2 Blue band via the SkyView Virtual Observatory, which provides sufficient resolution to identify the host galaxy and nearby cluster members.

We used archival X-ray data obtained by Chandra with the ACIS-I array (Obs ID: 15111).
We performed data reduction and imaging using CIAO version 4.16.0 \citep[][]{10.1117/12.671760} with CALDB 4.11.0 \citep[][]{10.1117/12.672876}.
The observation data were reprocessed with the \texttt{chandra\_repro} procedure, and exposure-corrected image in the 0.5--7.0~keV band were created using \texttt{flux\_image}, as shown by the cyan-contour in figure~\ref{fig:f00}.

\section{Result}\label{sec:3}

\subsection{Total Intensity Maps}

\begin{figure*}[tp]
\begin{center}
\includegraphics[width=8cm]{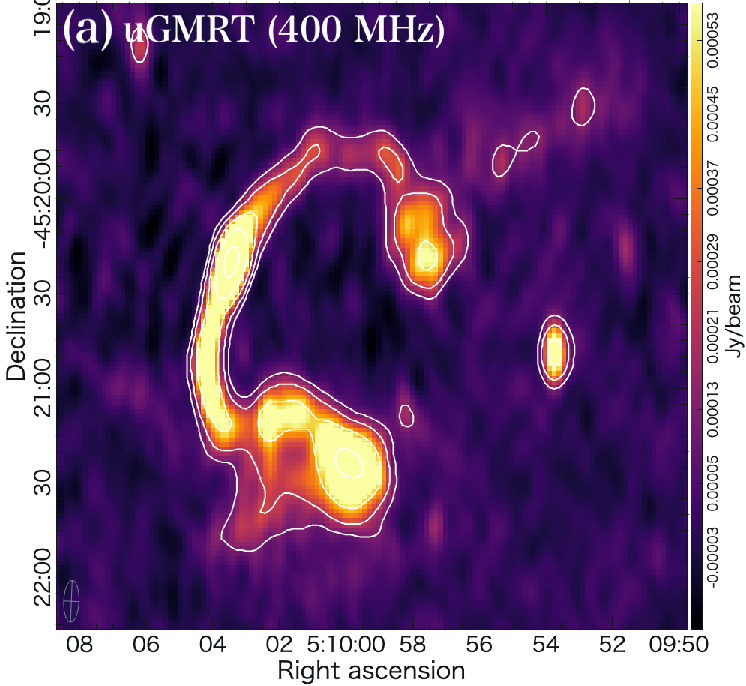}
\includegraphics[width=8cm]{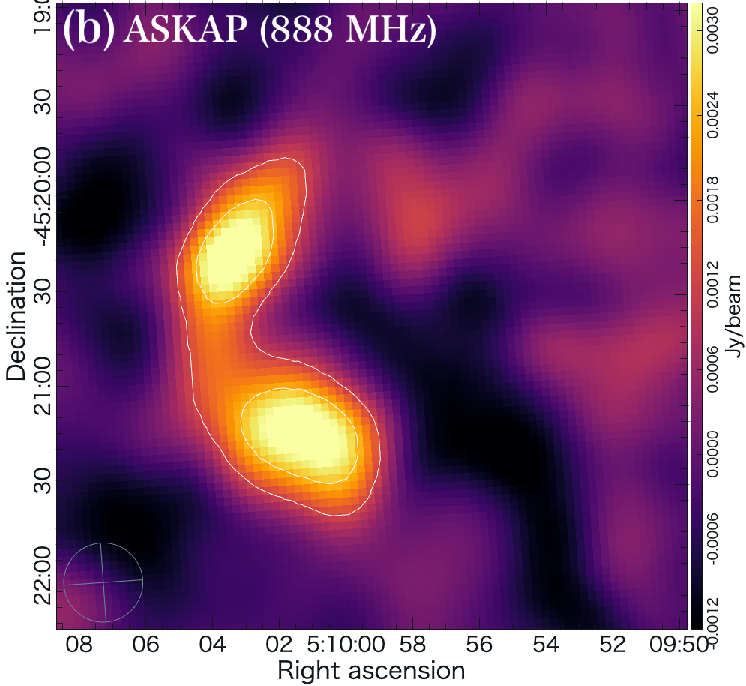}
\includegraphics[width=8cm]{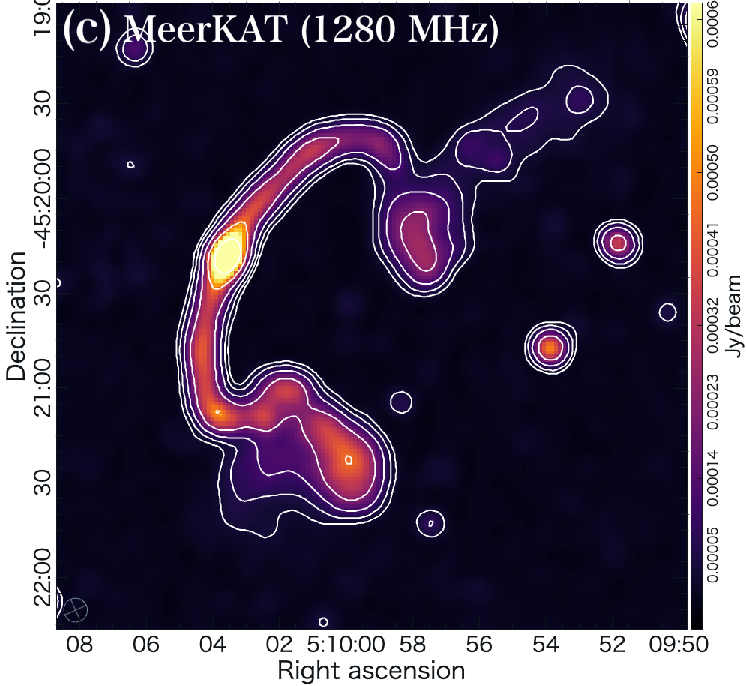}
\includegraphics[width=8cm]{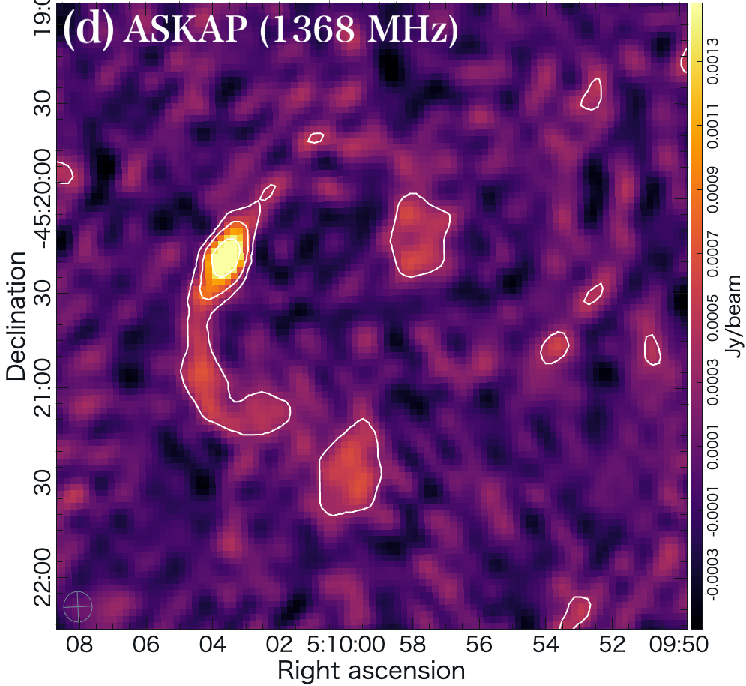}
\includegraphics[width=8cm]{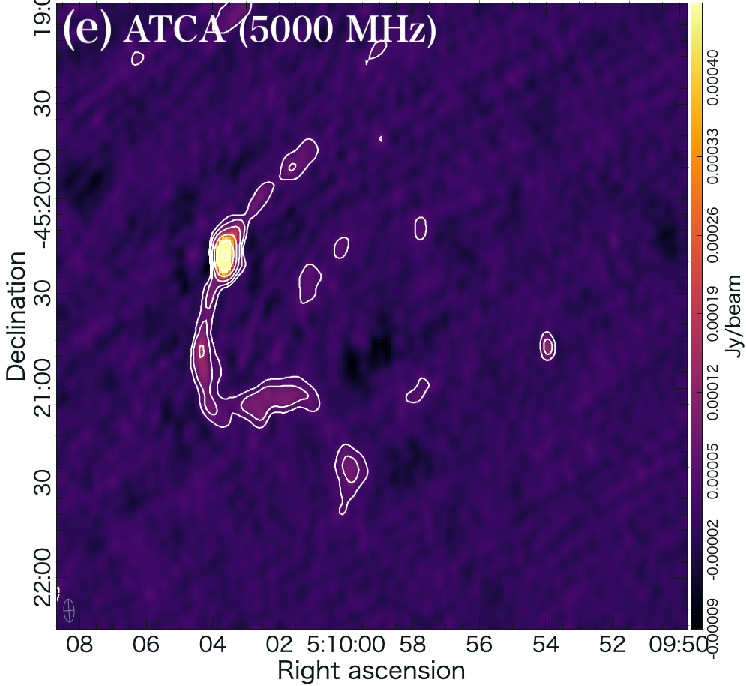}
\includegraphics[width=8cm]{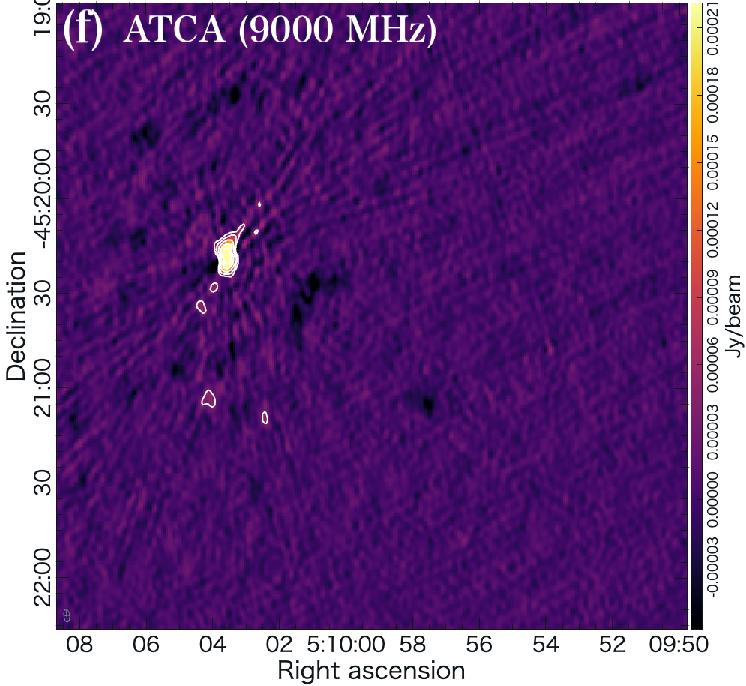}
\end{center}
\caption{Stokes I images of the Omega-Tail galaxy associated with Abell 3322, observed at multiple frequencies. Panels (a) to (f) correspond to the datasets listed in the same order as in Table~\ref{t01}. Each panel shows the total intensity map at a different frequency or resolution, revealing the detailed morphology of the radio jets and tails. The contours are drawn at $[3, 6, 12, 24, 48]\,\sigma_{\rm rms}$ levels, where $\sigma_{\rm rms}$ is the local root-mean-square noise measured in each map (see Table~\ref{tab:2} for the $\sigma_{\rm rms}$ values used in each dataset). The HT structure is clearly seen bending toward the south, likely due to ram pressure from the intracluster medium. The highest-frequency image (bottom right) shows compact core-like emission, while the lower-frequency images reveal extended, diffuse tails. These maps provide the basis for subsequent spectral index and spectral aging analyses.\\ 
{Alt text: Six panels of Stokes I images of the Omega-Tail galaxy showing radio jets and tails at different frequencies. The head-tail bends south, with a compact core visible at high frequency. }
}
\label{f01}
\end{figure*}

The total intensity maps of the Omega-tail galaxy, including archival data, are shown in Figure \ref{f01}. We discovered a unique Omega-shaped structure, which is not seen in typical WAT galaxies. Therefore, we call this head-tail galaxy associated with Abell 3322 as the Omega-tail galaxy. The Omega-tail consists of four parts: the core, two jets, two tails, and two splashes, described below.

We detected the head of the Omega-tail galaxy, or the ``core'' of the AGN jets, with a high signal-to-noise ratio (SNR) (${\rm SNR} > 100$) at all of the wavelengths we showed. The head is well associated with an optical galaxy. We also detected the slightly curved but well-collimated twin ``jets'', toward the north and south directions from the core, with a good SNR (${\rm SNR} > 10$) for the uGMRT and MeerKAT images, where the jets are about three times fainter than the core. The size along the north-south direction is 2 arcmin or 400 kpc, and is independent of the observed frequency. As for ASKAP RACS-low and ATCA 5 GHz, only the southern jet was detected in the images with an insufficient signal-to-noise ratio. At 9 GHz, we marginally detected only the launch point of the southern jet. 

Below 1.3~GHz, we found that both jets suddenly change their directions to the west while keeping similar brightness. We call this ``tails'' and distinguish them from the jets described above to make our discussion clear. The transition region of the jet and tail seems \FirstRevise{fainter} compared to the jets and tails. Interestingly, unlike the typical structure where the tail is deflected by up to 90 degrees, the tails of the Omega-tail galaxy bend more than 90 degrees. Meanwhile, there is a lobe structure at the terminal of the tails with an intensity peak like hot spots of jets. This structure is similar to what is observed in AGN radio lobes. Finally, we discovered with uGMRT and MeerKAT, which have good sensitivity, very faint structures beyond the terminals of the tails. Hereafter, we call this ``splashes'' structures. The splash structures are more than twice as faint as the jet components. \FirstRevise{Note that the Omega-shape represents only the projection onto the plane of the sky. The short splash on the south side might be due to it extending in the line of sight.
}
\subsection{Spectral Index Map}

\begin{figure}[tp]
\begin{center}
\includegraphics[width=7cm]{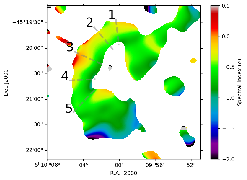}
\end{center}
\caption{The spectral index ($\alpha$) distribution of the Omega-tail galaxy, measured at an effective angular resolution of 15 arcseconds. The numbers from 1 to 5 and the gray dashed lines indicate the labels and positions of the slices used in the later discussion.\\
{Alt text: Spectral index ($\alpha$) map of the Omega-tail galaxy at 15 arcseconds resolution. Labels 1 to 5 mark slice positions shown by gray dashed lines.}
}
\label{f02}
\end{figure}

We derived the spectral index ($\alpha$ of $I \propto \nu^{\alpha}$) map using the total intensity maps. We used the uGMRT, MeerKAT, and ATCA data, while the RACS data were not used because MeerKAT data have a better noise level than RACS at a similar frequency. We divided the uGMRT data into 6 channels, MeerKAT data into 12 channels, and the ATCA data into 4 channels. The two lower frequency channels of the GMRT are not used because of their large beam size and poor data quality due to RFI flagging. We convolved them to a circular beam size of 15~arcseconds, which is about the major beam size of the GMRT's synthesized beam, and adjusted the pixel size and the pixel number of the maps. After that, we calculated the spectral index pixel by pixel by fitting a single power-law (SPL) function to the 20 frequency channels. The fitting was weighted by the inverse of each image's rms noise level. We can see a component that is dropped from the power law at high frequencies. We discuss this deviation from an SPL in detail in the next section.

The resultant spectral index map is shown in figure \ref{f02}. We found that the core region has the flattest index, approximately $-0.30$, and the spectrum continuously becomes steeper with increasing distance from the core, around $-0.56$ for the jets, around $-0.73$ for the tails, and around $-0.93$ for the splash, as the average value for the pixels within each part.

The steepening of the spectrum is also seen in the direction from the eastern rim to the western rim of the jets, i.e., in the direction perpendicular to the jet propagation (for further details, see Section 4.4).

\section{Discussion}\label{sec:4}

Our results suggest that the Omega-tail galaxy can be classified as a WAT. However, its very unique shape, resembling an Omega structure or splash structure, \FirstRevise{cannot be directly compared with typical WAT galaxies.} In this section, we further examine the radio characteristics of each structure and discuss how they have formed.

\subsection{Doppler Boost}

Figure \ref{f01} indicates that the southern jet tends to be brighter than the northern jet. We discuss here the possibility that this difference can be attributed to Doppler boosting, which occurs when the jet's propagation direction is not perfectly perpendicular to the line of sight. Using the angle between the line of sight and the jet direction, $\theta$ (where $0^\circ$ corresponds to a pole-on view), the relativistic beta value $\beta$, and the spectral index $\alpha$ of the jet, the brightness ratio R between the southern and northern jets can be expressed as follows \citep{1997MNRAS.288L...1H}:
\begin{equation}
    R = \Bigl( \frac{1 + \beta\ {\rm cos}\theta}{1 - \beta\ {\rm cos}\theta}\Bigl) ^{2 - \alpha}.
\label{equ1}
\end{equation}
The observed flux ratio of $R \sim 2$ implies a jet velocity of at least $\beta \sim 0.14$, even in the most extreme case of a pole-on view. This indicates that the jets are moving at sub-relativistic speeds. Such velocities are consistent with the general understanding that AGN jets, though initially relativistic at the core, typically decelerate to sub-relativistic speeds on kpc scales \citep{2009MNRAS.395..301W}. Based on this, it is natural to attribute the observed synchrotron emission to the sub-relativistic jets themselves. Conversely, if the emission were to originate from the surrounding shocked medium rather than the jets, one would expect little or no Doppler boosting, and the observed brightness asymmetry would be difficult to explain. This supports our interpretation that the emission is more plausibly from the jets. We note, however, that the pole-on geometry assumed above is an extreme case and not consistent with the observed morphology. Assuming that the emission does originate from the jets, a typical scenario for Fanaroff-Riley I (FRI) galaxies, the jet velocity would be approximately $\beta \sim 0.5$ \citep{2004MNRAS.351..727A}. In this case, reproducing the observed flux ratio of $R \sim 2$ would require the inclination angle to be $\theta = 74^\circ$. This inclination angle of $74^\circ$ is consistent with the implicit assumption in WAT and NAT sources that the jets are observed nearly side-on.

\subsection{Spectral Aging}

Cosmic-ray electrons are known to lose energy mainly through synchrotron radiation and inverse Compton scattering. At centimeter wavelength, higher-energy cosmic-ray electrons lose energy more rapidly than lower-energy cosmic-ray electrons. This results in a time evolution of the shape of the radio spectrum, also known as spectral aging. The spectral aging is commonly observed in head-tail galaxies and FR galaxies \citep{2013MNRAS.435.3353H}. The spectral aging is useful for estimating the age of radio structures. It also enables us to diagnose the potential of (re-)acceleration of cosmic rays inside the radio structure, providing physical insights into jet dynamics and interactions with the ICM.

The lifetime of the electrons emitting synchrotron radiation (Lorentz factor $\gamma \sim 10^4$) is given by the following equation \citep{2019SSRv..215...16V}.
\begin{equation}
    t_{\rm life~time} [{\rm yr}] = 3.2 \times 10^{10} \frac{B^{1/2}}{B^2+B_{\rm CMB}^2}[(1+z)\nu_{\rm obs}]^{-1/2}
\label{equ2}
\end{equation}
Here, $ B $ is the magnetic field strength in the emitting region in units of $\mu$G, $ B_{\rm CMB} $ is the equivalent magnetic field strength of the cosmic microwave background (CMB), and $ \nu_{\rm obs} $ is the observed frequency in MHz. At the redshift of Abell 3322, $ B_{\rm CMB} $ is 4.68~$\mu$G. If we assume energy equipartition between cosmic-ray electrons and the magnetic field, the average magnetic field strength of the Omega-tail galaxy is about 4~$\mu$G. Using these values, we obtain $ t_{\rm life~time} = 77 $~Myr at $ \nu = 400 $~MHz and $ t_{\rm life~time} = 24 $~Myr at $ \nu = 4000 $~MHz. These timescales are comparable to or less than the structure formation timescale of WAT galaxies. Therefore, energy loss is not negligible, and spectral aging could be seen in the Omega-tail galaxy. \FirstRevise{In particular, the non-detection of the northern emission in high frequency could be explained by spectral aging.}

\subsubsection{Model fitting}

To quantify the aging condition, we fitted the spectra using standard synchrotron spectrum models. Four different models were used for the fitting: a simple SPL model, as well as the so-called KP, JP, and CI models. 
The KP model \citep{1962SvA.....6..317K, 1970ranp.book.....P} assumes a single injection of electrons with constant pitch angles, leading to a high-frequency spectral steepening proportional to $\alpha_{\rm inj} - 1$. In contrast, the JP model \citep{1973A&A....26..423J} incorporates rapid pitch angle isotropization, resulting in an exponential cutoff beyond the break frequency. The CI model \citep{1994A&A...285...27K} considers continuous injection of electrons, producing a spectrum that remains flat at low frequencies and steepens by $\alpha_{\rm inj} - 0.5$ in spectral index beyond the break frequency. We used \texttt{synchrofit} \citep{2018MNRAS.476.2522T} for the KP, JP, and CI model fitting. 

\begin{figure}[tp]
\begin{center}
\includegraphics[width=7cm]{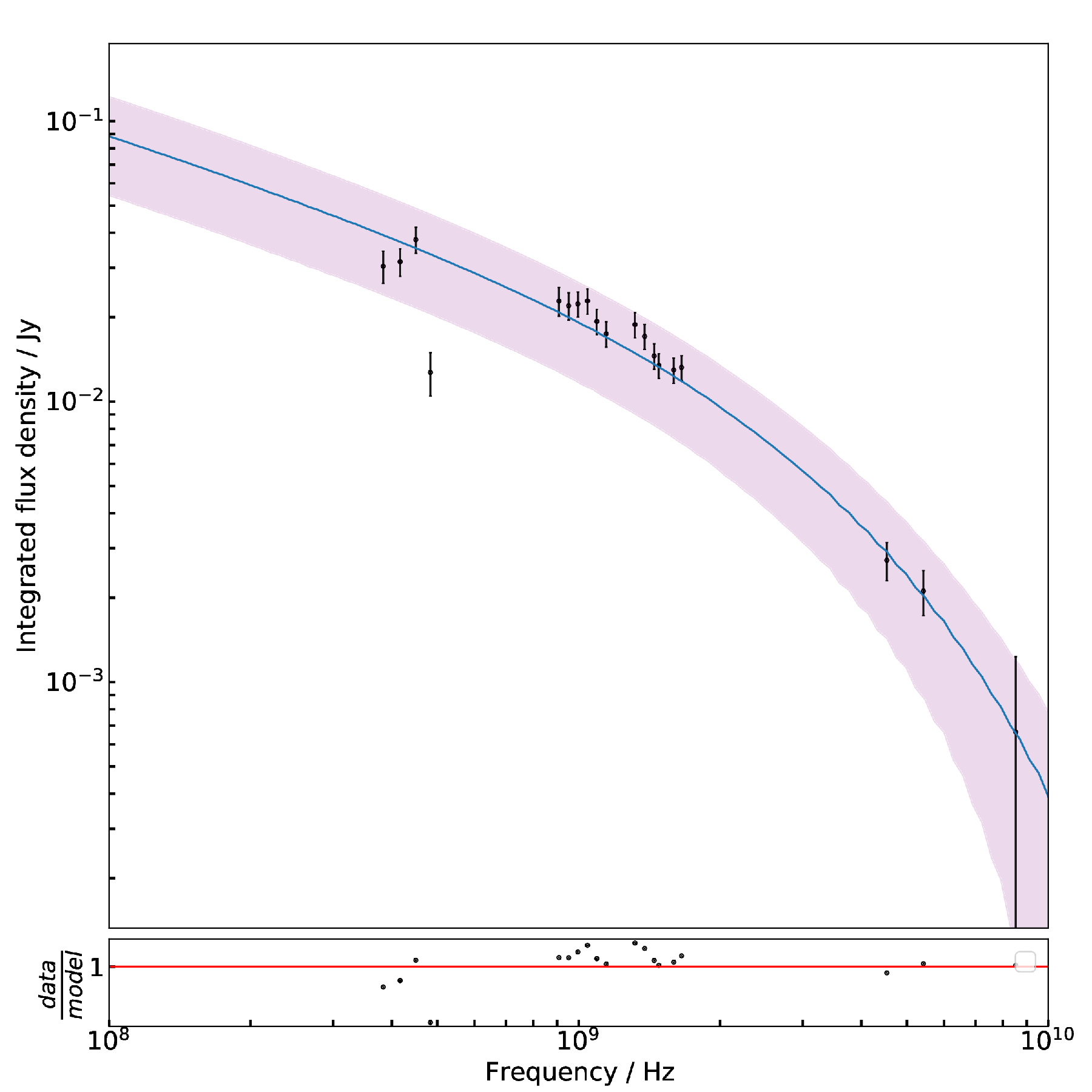}
\end{center}
\caption{The black points represent the observed data. The error bars were calculated as $\sqrt{\left( \sigma_{\rm rms} \sqrt{N_{\rm b}} \right)^2 + \left( \sigma_{\rm abs} F_\nu \right)^2}$, where $\sigma_{\rm rms}$ is the image noise, $N_{\rm b}$ is the number of beams within the total flux measurement region, and $\sigma_{\rm abs} = 0.1$ represents a 10\% uncertainty in the absolute flux scale. The solid line shows the best-fit curve obtained using `\texttt{synchrofit}' with the CI (continuous injection) model.\\
{Alt text: This figure shows the observed spectral data with uncertainties and the best-fit continuous injection (CI) model curve.}
} 
\label{fig:CI spectrum of the entire omega tail}
\end{figure}

Figure~\ref{fig:CI spectrum of the entire omega tail} shows the flux density spectrum, integrated within a circular region that encloses the entire Omega-tail structure, centered at (RA$_{\rm J2000}$, Dec$_{\rm J2000}$) = (05h09m59.5s, $-$45d20m40.0s), with a radius of 90~arcsec. A CI model was fitted to the spectrum, resulting in an injection spectral index of $\alpha_{\rm inj} = -0.5$, a spectral age of $38$~Myr, and an active duration of $4$~Myr. These results suggest that energy injection from the core responsible for forming the structure occurred only during an early phase. The derived total spectral age is significantly shorter than the dynamical timescale of the head-tail galaxy structure, indicating that re-acceleration likely plays a role in sustaining the observed emission.

\begin{figure*}[tp]
\begin{center}
\FigureFile(\linewidth,\linewidth){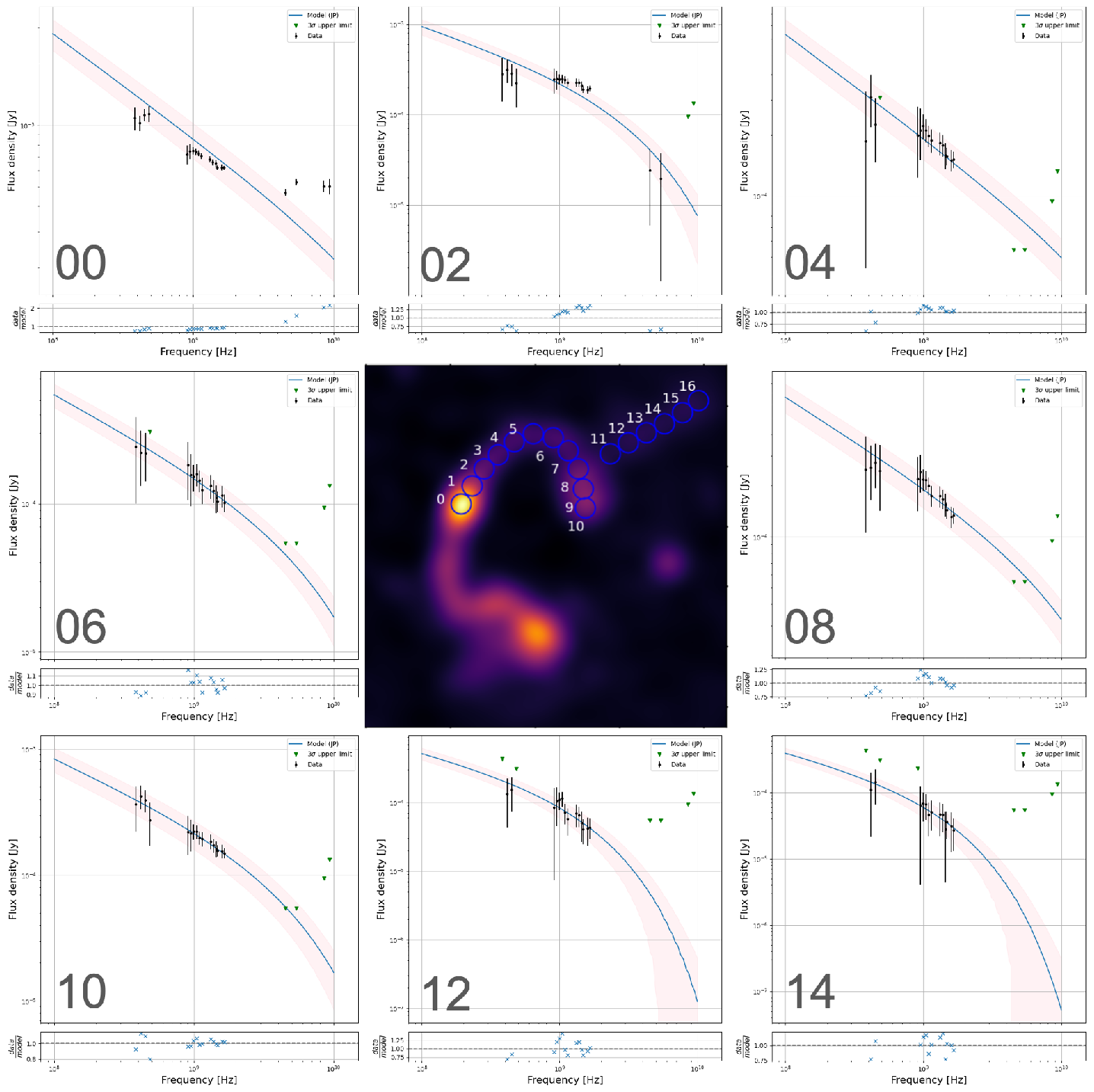}
\end{center}
\caption{The model fitting results for the northern component. The central panel shows the smoothed radio intensity map (MeerKAT), with circular markers indicating the regions from which the spectral fitting data were extracted. Surrounding the central image, the individual radio spectra corresponding to each extraction region are shown. The inset at the lower left corner illustrates the relative position numbers of circular markers. \FirstRevise{The black points represent the data with error bars. The green triangles indicate the upper limits for non-detected data, corresponding to 3$\sigma_{\rm rms}$. Note that these upper-limit data were not used in the fitting.}
\\
{Alt text: Model fitting results for the northern component. Central MeerKAT map shows extraction regions with circles. Surrounding panels display radio spectra for each region. Position numbers inset included.}
}
\label{synchrofit_north}
\end{figure*}

\begin{figure*}[tp]
\begin{center}
\FigureFile(\linewidth,\linewidth){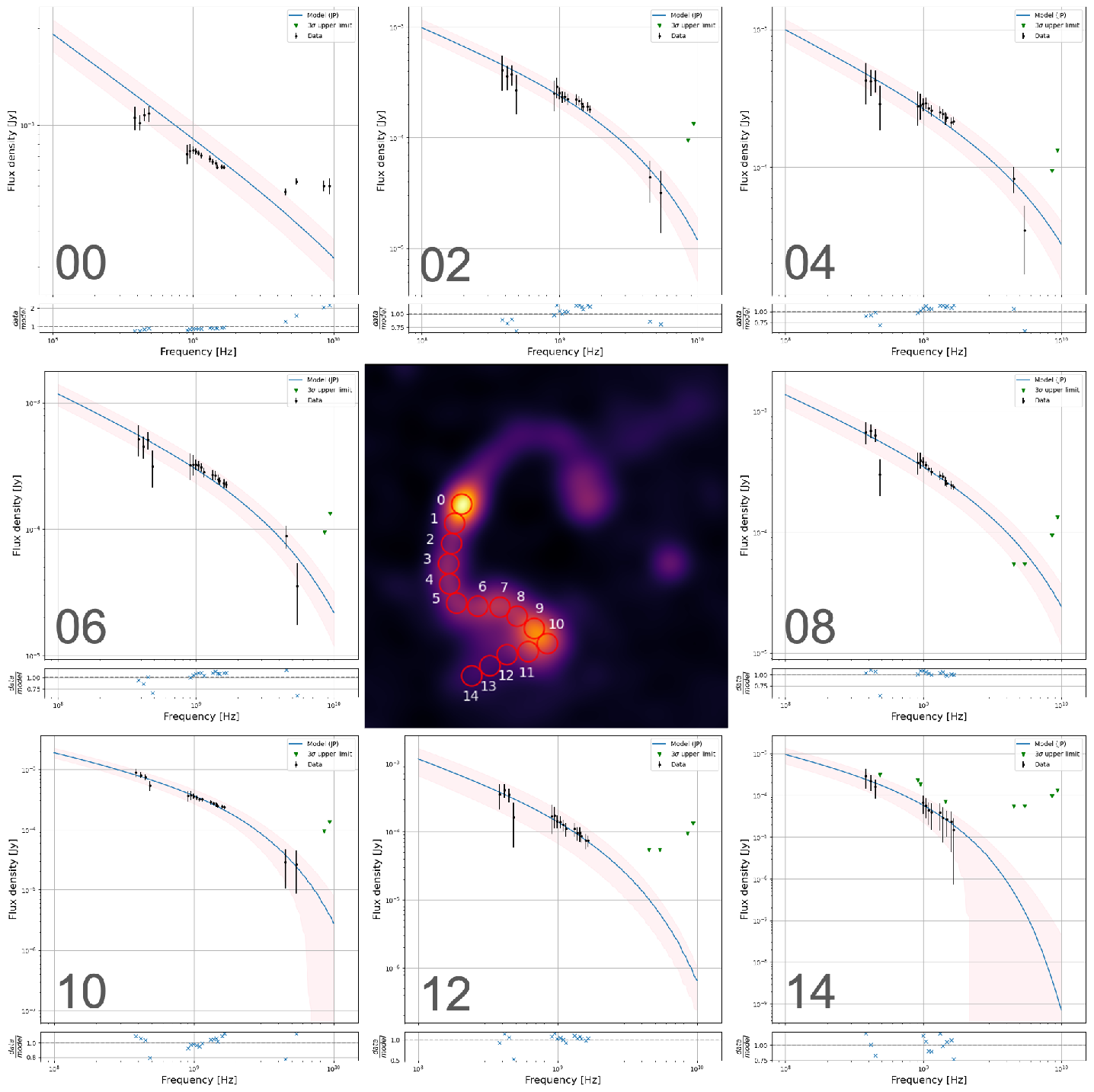}
\end{center}
\caption{The same as figure~\ref{synchrofit_north}, but for the southern component.\\
{Alt text: The same as figure~\ref{synchrofit_north}, but for the southern component.}
}
\label{synchrofit_south}
\end{figure*}

We examine the spatial variation of the aging spectrum along the Omega structure. The noise of the flux density $\sigma_{\rm flux}$ is determined using the intensity-based noise level $\sigma_{\rm rms}$, averaged over the integration area used for flux density calculation. $\sigma_{\rm flux}$ is given by
\begin{equation}
    \sigma_{\rm flux} = \sigma_{\rm rms} \sqrt{N_{\rm b}} ,
\label{equ3}
\end{equation}
where $N_{\rm b}$ is the number of beams in the flux density calculation area. The corresponding spectral fitting results are shown in figures~\ref{synchrofit_north} and \ref{synchrofit_south}, and the fitted parameters are summarized in table~\ref{tab:2}. \FirstRevise{Note that these upper-limit data were not used in the fitting.} Overall, the northern side tends to have younger spectral ages than the southern side. While this may reflect the effect of vertical shocks, which will be discussed later (in section 4.4), it is also possible that the fitting is biased due to the lack of high-frequency data, particularly in the northern region. As seen in figures~\ref{f01}, \ref{synchrofit_north}, and \ref{synchrofit_south}, the high-frequency (i.e., ATCA) data are notably faint on the northern side. Therefore, the following discussion in this subsection focuses only on the southern component. \FirstRevise{To further assess this effect in the southern component, we compared the fitting results with three times the noise level ($3\sigma_{\rm flux}$) as an upper limit for frequencies where the emission was not detected. As a result, for all regions except No. 8, the fitted lines are expected to lie below these upper limits, suggesting that our method is largely reliable.
For No. 8, the upper limit is slightly below the fitted line; however, since it is nearly comparable to the fitted value, we conclude that this will not significantly affect any further analysis or interpretation. With next-generation high-sensitivity and high-resolution observatories like the SKAO, these details can be studied more thoroughly.}

The reduced $\chi^2$ values are summarized in table~\ref{tab:3}. We found that the reduced $\chi^2$ is the smallest with the CI and JP models for the spots except No. 00, 03, 04, and 05, where the SPL model is the best fitted. For spots No.12 and beyond, the reduced $\chi^2$ is below unity for all models. These spots exhibit faint emission, and the error relative to the data values is larger and overestimated, consequently making the reduced $\chi^2$ below unity. Although the spots were initially selected based on a 3$\sigma_{\rm rms}$ significance, the spots No.12 and beyond may be too faint to be suitable for a precise error estimate and reliable fitting. Given that the JP model generally yields the lowest reduced $\chi^2$ values among the tested models, and assuming the absence of continuous injection, we adopt the JP model for further analysis. 

\begin{figure}[tp]
\begin{center}
\FigureFile(\linewidth,\linewidth){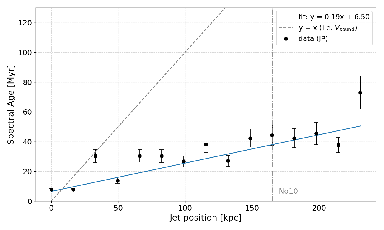}
\end{center}
\caption{Spectral age variation along the Southern Omega tail. The black dots represent the data derived from spectral fitting with the JP model, and the blue solid line shows the best-fit linear trend. For reference, the expected spectral age evolution under a simple synchrotron cooling scenario, assuming the plasma propagates at the sound speed ($1000~{\rm km,s^{-1}}$), is indicated by the grey dashed line. The vertical dot-dashed line marks the position of No.10, which is interpreted as a potential hotspot.\\
{Alt text: Variation of spectral age along the Southern Omega tail shown by JP model data (black dots) and linear fit (blue line). Hotspot at position No.10 marked.}
}\label{spectral age}
\end{figure}

Figure~\ref{spectral age} shows the profile of the spectral age along the Omega structure. The spectral age appears to increase monotonically with distance from the core. A linear fit to the entire profile yields a slope of $0.2$~Myr~kpc$^{-1}$. If we consider only simple radiative cooling, the spectral age should directly reflect the travel time from the core to the given position. Under the assumption that the gas moves at the sound speed in the ICM, approximately 1000~km~s$^{-1}$, the spectral age is expected to evolve along the grey dashed line shown in figure \ref{spectral age}. We found that the observed spectral age grows much slower than the sound speed prediction, suggesting that an anti-aging effect, such as re-acceleration and/or recharging of cosmic rays, is counteracting spectral aging. 

\clearpage
\begin{sidewaystable*}[H]
  \centering
  \caption{Summary of spectral‐fitting results for the JP, KP, and CI models, reporting the fitted spectral index, break frequency (in log scale), spectral age, and CI‐model active duration for both the northern and southern jets.}
  \label{tab:2}
  \begin{tabular}{ccrrrrrrrrrr}
    \toprule
    & & \multicolumn{3}{c}{JP} & \multicolumn{3}{c}{KP} & \multicolumn{4}{c}{CI} \\
    \cmidrule(lr){3-5} \cmidrule(lr){6-8} \cmidrule(lr){9-12}
    Group & Pos & Index($s$) & Break freq & Age & Index($s$) & Break freq & Age & Index($s$) & Break freq & Age & Active \\
    - & - & $\alpha = -(s-1)/2$ & [Hz in log] & [Myr] & $\alpha = -(s-1)/2$ & [Hz in log] & [Myr] & $\alpha = -(s-1)/2$ & [Hz in log] & [Myr] & [Myr] \\
    \midrule
    \multirow{17}{*}{North} & 00 & $2.01\pm0.02$ & $11.00\pm0.06$ & $7.85\pm0.52$ & $2.01\pm0.02$ & $10.99\pm0.05$ & $3.53\pm0.45$ & $2.01\pm0.02$ & $11.00\pm0.06$ & $7.85\pm0.51$ & 0.51 \\
    & 01 & $2.01\pm0.02$ & $11.00\pm0.06$ & $7.85\pm0.52$ & $2.01\pm0.02$ & $10.98\pm0.06$ & $3.55\pm0.51$ & $2.01\pm0.02$ & $11.00\pm0.05$ & $7.88\pm0.49$ & 0.49 \\
    & 02 & $2.01\pm0.02$ & $9.65\pm0.12$ & $37.10\pm5.10$ & $2.01\pm0.02$ & $9.46\pm0.12$ & $20.45\pm6.25$ & $2.01\pm0.02$ & $9.65\pm0.12$ & $37.10\pm4.27$ & 5.10 \\
    & 03 & $2.01\pm0.02$ & $10.91\pm0.06$ & $8.74\pm0.61$ & $2.01\pm0.02$ & $10.67\pm0.19$ & $5.07\pm2.51$ & $2.01\pm0.02$ & $10.91\pm0.06$ & $8.74\pm1.01$ & 0.61 \\
    & 04 & $2.01\pm0.02$ & $10.36\pm0.17$ & $16.31\pm3.28$ & $2.01\pm0.02$ & $9.97\pm0.15$ & $11.37\pm4.34$ & $2.01\pm0.02$ & $10.36\pm0.17$ & $16.31\pm1.88$ & 3.28 \\
    & 05 & $2.01\pm0.02$ & $10.98\pm0.05$ & $7.99\pm0.50$ & $2.01\pm0.02$ & $10.96\pm0.05$ & $3.63\pm0.48$ & $2.01\pm0.02$ & $10.97\pm0.07$ & $8.08\pm0.63$ & 0.63 \\
    & 06 & $2.01\pm0.02$ & $10.85\pm0.08$ & $9.31\pm0.85$ & $2.01\pm0.02$ & $10.74\pm0.12$ & $4.69\pm1.49$ & $2.01\pm0.02$ & $10.85\pm0.08$ & $9.31\pm1.07$ & 0.85 \\
    & 07 & $2.01\pm0.02$ & $9.88\pm0.12$ & $28.35\pm3.86$ & $2.01\pm0.02$ & $9.61\pm0.13$ & $17.24\pm5.72$ & $2.01\pm0.02$ & $9.88\pm0.12$ & $28.35\pm3.27$ & 3.86 \\
    & 08 & $2.01\pm0.02$ & $10.91\pm0.07$ & $8.68\pm0.70$ & $2.01\pm0.02$ & $10.61\pm0.12$ & $5.45\pm1.71$ & $2.01\pm0.02$ & $10.75\pm0.12$ & $10.41\pm5.09$ & 1.41 \\
    & 09 & $2.01\pm0.02$ & $10.92\pm0.04$ & $8.56\pm0.40$ & $2.01\pm0.02$ & $10.89\pm0.06$ & $3.95\pm0.61$ & $2.01\pm0.02$ & $10.88\pm0.06$ & $8.96\pm0.65$ & 0.65 \\
    & 10 & $2.01\pm0.02$ & $9.83\pm0.13$ & $30.36\pm4.48$ & $2.01\pm0.02$ & $9.62\pm0.12$ & $17.03\pm5.39$ & $2.01\pm0.02$ & $9.84\pm0.12$ & $29.99\pm3.44$ & 4.26 \\
    & 11 & $2.01\pm0.02$ & $9.54\pm0.12$ & $42.36\pm5.72$ & $2.01\pm0.02$ & $9.28\pm0.15$ & $25.23\pm9.78$ & $2.01\pm0.02$ & $9.41\pm0.13$ & $48.83\pm7.23$ & 7.23 \\
    & 12 & $2.01\pm0.02$ & $9.37\pm0.12$ & $51.17\pm7.00$ & $2.01\pm0.02$ & $9.18\pm0.12$ & $28.21\pm8.65$ & $2.01\pm0.02$ & $9.37\pm0.12$ & $51.17\pm7.00$ & 7.00 \\
    & 13 & $2.01\pm0.02$ & $9.41\pm0.13$ & $48.83\pm7.19$ & $2.01\pm0.02$ & $9.27\pm0.12$ & $25.44\pm8.14$ & $2.01\pm0.02$ & $9.41\pm0.13$ & $48.83\pm6.12$ & 7.20 \\
    & 14 & $2.08\pm0.02$ & $9.37\pm0.12$ & $51.17\pm7.10$ & $2.01\pm0.02$ & $9.06\pm0.12$ & $32.41\pm10.20$ & $2.01\pm0.02$ & $9.25\pm0.13$ & $58.98\pm8.73$ & 9.01 \\
    & 15 & $2.13\pm0.07$ & $9.41\pm0.13$ & $48.83\pm7.41$ & $2.01\pm0.02$ & $9.10\pm0.13$ & $31.00\pm10.40$ & $2.01\pm0.02$ & $8.96\pm0.12$ & $81.83\pm11.28$ & 11.28 \\
    & 16 & $2.01\pm0.02$ & $9.28\pm0.13$ & $56.60\pm8.43$ & $2.01\pm0.02$ & $9.06\pm0.12$ & $32.41\pm9.86$ & $2.01\pm0.02$ & $9.12\pm0.13$ & $68.13\pm9.89$ & 9.89 \\
    \midrule
    \multirow{15}{*}{South} & 00 & $2.01\pm0.02$ & $11.00\pm0.06$ & $7.85\pm0.52$ & $2.01\pm0.02$ & $10.99\pm0.05$ & $3.53\pm0.45$ & $2.01\pm0.02$ & $11.00\pm0.06$ & $7.85\pm0.51$ & 0.51 \\
    & 01 & $2.01\pm0.02$ & $11.00\pm0.05$ & $7.88\pm0.50$ & $2.01\pm0.02$ & $10.98\pm0.05$ & $3.58\pm0.46$ & $2.01\pm0.02$ & $10.99\pm0.05$ & $7.97\pm0.42$ & 0.42 \\
    & 02 & $2.01\pm0.02$ & $9.83\pm0.13$ & $30.36\pm4.40$ & $2.01\pm0.02$ & $9.64\pm0.12$ & $16.78\pm5.18$ & $2.01\pm0.02$ & $9.83\pm0.13$ & $30.36\pm3.50$ & 4.40 \\
    & 03 & $2.01\pm0.02$ & $10.32\pm0.12$ & $17.12\pm2.31$ & $2.01\pm0.02$ & $10.14\pm0.13$ & $9.35\pm3.12$ & $2.01\pm0.02$ & $10.06\pm0.12$ & $23.27\pm12.16$ & 3.13 \\
    & 04 & $2.01\pm0.02$ & $9.98\pm0.12$ & $25.49\pm3.56$ & $2.01\pm0.02$ & $9.76\pm0.12$ & $14.48\pm4.48$ & $2.01\pm0.02$ & $9.97\pm0.13$ & $25.56\pm2.93$ & 3.82 \\
    & 05 & $2.01\pm0.02$ & $9.94\pm0.12$ & $26.73\pm3.62$ & $2.01\pm0.02$ & $9.76\pm0.12$ & $14.48\pm4.40$ & $2.01\pm0.02$ & $9.94\pm0.12$ & $26.73\pm3.62$ & 3.62 \\
    & 06 & $2.01\pm0.02$ & $9.98\pm0.12$ & $25.49\pm3.54$ & $2.01\pm0.02$ & $9.76\pm0.12$ & $14.48\pm4.46$ & $2.01\pm0.02$ & $9.97\pm0.13$ & $25.56\pm2.93$ & 3.79 \\
    & 07 & $2.01\pm0.02$ & $9.63\pm0.13$ & $38.22\pm5.59$ & $2.01\pm0.02$ & $9.44\pm0.12$ & $21.12\pm6.59$ & $2.01\pm0.02$ & $9.63\pm0.13$ & $38.22\pm4.40$ & 5.59 \\
    & 08 & $2.01\pm0.05$ & $9.94\pm0.18$ & $26.76\pm5.46$ & $2.01\pm0.02$ & $9.74\pm0.12$ & $14.96\pm4.54$ & $2.04\pm0.02$ & $9.76\pm0.18$ & $32.54\pm16.87$ & 6.77 \\
    & 09 & $2.01\pm0.02$ & $9.49\pm0.13$ & $44.81\pm6.80$ & $2.01\pm0.02$ & $9.26\pm0.13$ & $25.75\pm8.50$ & $2.01\pm0.02$ & $9.49\pm0.13$ & $44.81\pm5.16$ & 6.80 \\
    & 10 & $2.01\pm0.02$ & $9.49\pm0.13$ & $44.81\pm6.95$ & $2.01\pm0.02$ & $9.26\pm0.13$ & $25.75\pm8.71$ & $2.01\pm0.02$ & $9.49\pm0.13$ & $44.81\pm5.31$ & 6.95 \\
    & 11 & $2.01\pm0.02$ & $9.54\pm0.13$ & $42.36\pm6.45$ & $2.01\pm0.02$ & $9.31\pm0.13$ & $24.34\pm8.18$ & $2.01\pm0.02$ & $9.54\pm0.13$ & $42.36\pm4.86$ & 6.45 \\
    & 12 & $2.31\pm0.07$ & $9.47\pm0.14$ & $45.46\pm7.37$ & $2.18\pm0.04$ & $9.16\pm0.13$ & $28.89\pm9.43$ & $2.22\pm0.07$ & $9.23\pm0.18$ & $60.51\pm22.88$ & 12.75 \\
    & 13 & $2.99\pm0.04$ & $9.73\pm0.16$ & $34.03\pm6.14$ & $2.97\pm0.01$ & $9.46\pm0.12$ & $20.52\pm6.62$ & $2.99\pm0.02$ & $8.73\pm0.12$ & $107.71\pm103.36$ & 15.31 \\
    & 14 & $2.39\pm0.04$ & $9.06\pm0.13$ & $72.93\pm10.96$ & $2.46\pm0.07$ & $8.87\pm0.14$ & $40.31\pm14.54$ & $2.19\pm0.04$ & $8.97\pm0.15$ & $80.84\pm10.45$ & 14.34 \\
    \bottomrule
  \end{tabular}
\end{sidewaystable*}

\clearpage

\begin{table}[htbp]
\centering
\caption{Reduced chi-squared values for each spectral model at each region.}\label{tab:3}
\begin{tabular}{c|rrrr}
\hline
No & SPL & JP & KP & CI \\
\hline
0  & 10.8 & 81.9 & 73.9 & 66.3 \\
1  &  2.1 &  2.4 &  2.2 &  2.1 \\
2  &  4.4 &  3.6 &  3.8 &  3.6 \\
3  &  3.7 &  4.4 &  4.5 &  4.5 \\
4  &  4.8 &  5.0 &  5.1 &  5.0 \\
5  &  9.0 & 11.4 & 11.7 & 11.4 \\
6  &  5.3 &  4.5 &  4.6 &  4.5 \\
7  &  8.2 &  6.9 &  7.5 &  6.9 \\
8  &  3.2 &  3.1 &  3.2 &  3.1 \\
9  & 11.2 &  6.7 &  8.8 &  6.7 \\
10 &  5.3 &  2.4 &  3.0 &  2.4 \\
11 &  3.8 &  1.5 &  1.8 &  1.5 \\
12 &  1.0 &  0.9 &  0.9 &  0.9 \\
13 &  0.6 &  0.5 &  0.5 &  0.5 \\
14 &  0.4 &  0.3 &  0.3 &  0.2 \\
\hline
\end{tabular}
\end{table}

\subsubsection{Comparison with simulated data}

\begin{figure}[tp]
\begin{center}
\FigureFile(\linewidth,\linewidth){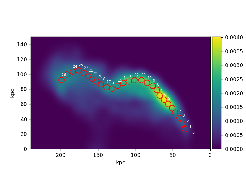}
\FigureFile(\linewidth,\linewidth){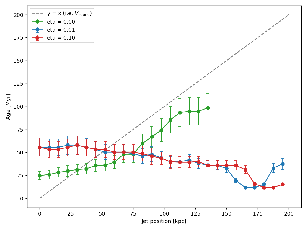}
\end{center}
\caption{Pseudo-radio observations derived from head-tail galaxy simulation data \citep{2023ApJ...951...76O}.  (\textbf{Top}) Simulated radio image at 500~MHz assuming a re-acceleration efficiency of \FirstRevise{0.1}.  (\textbf{Bottom}) Variation of the spectral age along the jet propagation axis. The regions used to calculate the flux density differ for each re-acceleration efficiency model, but in general, they trace the bright central part of the jet/tail structure. For example, in the case of the \FirstRevise{$\eta = 0.1$} model, the region is indicated by the red circles in the top panel. The grey dashed line indicates the same reference sound-speed track ($1000,{\rm km,s^{-1}}$) as shown in figure~\ref{spectral age}. The green, blue, and red dotted lines correspond to the spectral ages inferred using the JP model applied to the simulation data with re-acceleration efficiencies $\eta$ of $\eta = 0.00$ (no re-acceleration), $\eta = 0.01$, and $\eta = 0.10$, respectively. \\
{Alt text: Simulated 500 MHz radio image (top) and spectral age variation (bottom) of head-tail galaxy jets with different re-acceleration efficiencies shown by colored lines.}
}
\label{Ohmura age}
\end{figure}

We further test the observed spectral aging and anti-aging trend using MHD simulation data from \citet{2023ApJ...951...76O}. In \citet{2023ApJ...951...76O}, the formation of a head-tail galaxy was simulated under the influence of ram pressure resulting from the motion of the host galaxy through the ICM. The simulation tracks the evolution of cosmic-ray electron energies, incorporating Fermi II-type turbulent re-acceleration, in order to model the spatial distribution of synchrotron age from radio observations. Three different acceleration models were considered in the simulation, with the re-acceleration efficiency parameter, $\eta=0$ (no re-acceleration model), \FirstRevise{$0.01$} (weak re-acceleration model), and \FirstRevise{$0.1$} (strong re-acceleration model). See \citet{2023ApJ...951...76O} for details of the simulation.

We use the simulated data from 100~MHz to 10~GHz, containing 20 frequency channels for the spectral fitting. We applied the spectral model fitting in the same way as the one for the observation data. The upper panel of figure \ref{Ohmura age} shows the image of the simulated tail with $\eta=0.01$. The circles on the tail indicate the areas considered for the spectral-age profile along the jet. The lower panel of figure \ref{Ohmura age} shows the variation of spectral age along the tail direction with the three different acceleration models. We confirmed that spectral aging follows the sound speed prediction for the case of no re-acceleration model, while the two re-acceleration models indicate nearly constant spectral age along the tail. Comparing these simulations with our observations, it is likely that the re-acceleration with an efficiency of less than 0.01 is occurring in the Omega-tail galaxy. 

\subsection{Cosmic Ray Reacceleration}

Our discussion in the previous subsection suggests that the cosmic-ray electrons inside the jets are being re-accelerated. In this subsection, we discuss the re-acceleration mechanism in the Omega-tail galaxy.

\begin{figure}[tp]
\begin{center}
\FigureFile(\linewidth,\linewidth){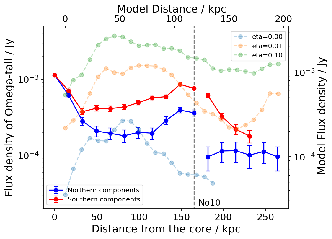}
\FigureFile(\linewidth,\linewidth){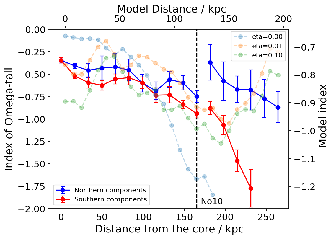}
\end{center}
\caption{Radio flux density profile along the Omega-tail at 400 MHz observed with GMRT (top), and spectral index profile along the Omega structure (bottom). In both panels, the bold red and blue lines represent the southern and northern components of the Omega-tail galaxy, respectively. The semi-transparent lines represent the data calculated from the model \citep{2023ApJ...951...76O}. For the axes, the observed data correspond to the left and bottom axes, while the simulation data correspond to the right and top axes.\\
{Alt text: Radio flux density (400 MHz) and spectral index profiles along the Omega-tail. Southern and northern components shown in bold red and blue; model data in semi-transparent lines.}
}
\label{f07}
\end{figure}

Firstly, the re-acceleration of cosmic-ray electrons can influence the brightness of synchrotron emission. The upper panel of Figure \ref{f07} shows the variations in brightness along the Omega structure \FirstRevise{, overlaid with the same profile derived from the simulation.} Both the northern and southern components exhibit similar trends: the core is bright, the brightness increases along the jet with some regions remaining flat, and there is a discontinuity at spot No.10 in the Northern component. These features, including the presence of hotspots and an overall increasing trend, can be explained by the occurrence of weak re-acceleration in both jets when compared with simulations \citep{2023ApJ...951...76O}.

Secondly, the variation of the spectral index serves as a crucial indicator of the presence of a re-acceleration process. The lower panel of figure~\ref{f07} shows the spectral index profile along the Omega structure \FirstRevise{, overlaid with the simulation.} Here, we adopt the spectral index fitted with the SPL model for a comparison with the simulation results of \citet{2023ApJ...951...76O}. The profile, except for the discontinuous bend at No. 10, shows a monotonic decrease slower than pure synchrotron cooling/aging. This trend is commonly observed in other head-tail galaxies, supporting re-acceleration \citep[e.g.,][]{2024MNRAS.528..141L,2021MNRAS.508.5326M,2020MNRAS.493.3811S}. According to numerical simulations, the aging rate varies with the re-acceleration efficiency \citep{2023ApJ...951...76O}. In the Omega-tail galaxy, the spectral index, which was $\alpha = -0.4$ at the jet core, has shifted to $\alpha = -0.7$ on the northern side and $\alpha = -0.9$ on the southern side at a distance of 150 kpc. This difference of approximately $\Delta \alpha = -0.4$ is consistent with the case of an efficiency \FirstRevise{$\eta \sim 0.01$ and $0.1$} re-acceleration, as suggested by \citet{2023ApJ...951...76O}. 

\FirstRevise{In addition, the simulation shows that the spectrum becomes softer in the downstream region of the tail (beyond 150 kpc). This is because older electrons accumulate there and have reached the re-acceleration timescale \citep[see][Eq. 6]{2023ApJ...951...76O}. As can be seen in Figure \ref{Ohmura age}, along the jet in the simulation, acceleration and cooling are roughly balanced, or acceleration is slightly dominant. In the downstream region of the tail, the gas velocity decreases and electrons of various (actual) ages coexist. Among these, older electrons have experienced prolonged periods of acceleration-dominated conditions, which likely leads to the softening of the spectral shape.
These effects have not been directly confirmed in observations. This may indicate that the re-acceleration efficiency in the observed Omega tail is even weaker than that considered in the simulation of $\eta = 0.01$. Alternatively, the same behavior can be reproduced if turbulence dissipates more rapidly (i.e., if the dissipation scale is larger). Interestingly, this appears to be consistent with recent direct measurements of turbulent velocities in the Coma cluster with high spectral resolution by XRISM \citep{2025ApJ...985L..20X}.
}

Finally, from the overall trend of the spectral index profile, the southern jet tends to have a steeper spectrum compared to the northern jet. The beaming effect \FirstRevise{has no frequency dependence, and therefore} cannot explain this systematic difference of spectral indices, unless the inclination angle $\theta$ differs from north to south. A possibility of this systematic difference is that the electrons in the northern jet are less cooling and/or more re-accelerated due to additional re-acceleration sources. A candidate for the additional sources is that the northern jet exhibits a noticeable gradient of spectral index in the direction perpendicular to the jet propagation. We suggest that this gradient of the spectral index is caused by a shock front of the infalling head-tail galaxy toward the cluster center. This shock could form the systematic difference in the spectral index between the northern and southern jets. We discuss the perpendicular shock in the next subsection.
\subsection{Perpendicular Shock}

\begin{figure}[tp]
\begin{center}
\FigureFile(\linewidth,\linewidth){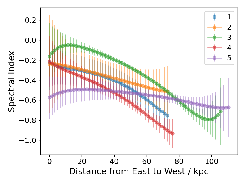}
\end{center}
\caption{Spectral index variation along the dashed line shown in figure~\ref{f02}. The horizontal axis represents the projected distance from east to west.\\
{Alt text: Spectral index variation along the dashed line shown in figure~\ref{f02}. The horizontal axis represents the projected distance from east to west.}
}\label{f08}
\end{figure}

Figure~\ref{f08} shows spectral index profiles measured perpendicular to the jet propagation axis. These profiles exhibit a gradient from the upstream (eastern) side to the downstream (western) side, suggesting influences of the ICM wind shock. In general, when fresh electrons are injected from the core, the flattest spectra is expected along the jet axis. This trend should hold even at lower spatial resolution. However, in the case of the Omega-tail galaxy, the observed distribution deviates from this expectation. 
We also derived the spectral index distribution of the simultaneously observed BCG using the same procedure. No such trend was found; instead, the flattest spectra were located at the center of the source, and the spectrum became progressively steeper toward the edges of the structure. Therefore, the observed gradient is unlikely to be an artifact caused by beam smearing. Although the origin of this gradient remains unclear, one possibility is that shock waves are generated ahead of the Omega-tail galaxy as it moves through the ICM, accelerating electrons in the process.

Indeed, oblique shocks induced by jet injection have also been found in simulations \citep{1992ApJ...393..631B, 2023ApJ...951...76O}. If this is the case, the spectrum is expected to be the flattest at the position closest to the shock wave, which is consistent with the case of the Omega-tail galaxy. The spectral index gradient of the Omega-tail galaxy is not perfectly parallel to the shape of the jet, which implies that the shock waves are oriented slightly to the northern side. This suggests that the Omega-tail galaxy is falling toward the cluster center with a small counter-clockwise motion, and the northern jet is facing the shock wave. This is broadly consistent with the fact that the northern jet has a less-aged spectrum compared to the southern jet, as discussed in the previous subsection. 

In a virialized galaxy cluster, it is expected that the velocity of the proper motion of a member galaxy is comparable to or less than the sound velocity of the ICM, so that the potential shock wave caused by the motion of the galaxy should be of order unity. In such a small Mach number shock, the standard diffusive shock acceleration \citep[DSA;][]{1987PhR...154....1B} does not efficiently accelerate cosmic rays. The simplest interpretation is that the Omega-tail galaxy possesses a peculiarly high velocity and thus the shock has a higher Mach number, though there is no supporting evidence for this hypothesis. An elongation of the X-ray surface brightness profile (see figure \ref{fig:f00}) rather suggests that a cluster merger, which possibly induced high velocity galaxies, happened in the past along the direction perpendicular to the expected infalling direction of the Omega-tail galaxy. This prediction will be studied more with a future high-sensitivity and high-resolution telescope such as the SKAO.

Another possibility of this spatial gradient of spectral index is that older electrons are accumulated within the $\Omega$-shaped structure of the jet, either through advection by backflowing plasma or by becoming trapped in Kelvin-Helmholtz (KH) vortices. Such features are also seen in MHD simulations (see figure 1 of \citealt{2023ApJ...951...76O}). However, considering the sub-relativistic motion of the jet, the aging effect would be minor along the direction perpendicular to the jet propagation.

\subsection{Origin of "Omega" structure}

The "Omega" structure in its shape of the observed head-tail galaxy is very unique and not usually seen in other head-tail galaxies. Generally speaking, the tail component could be formed by a wind from the ICM, i.e., ram pressure, when the galaxy moves into the cluster. But in theory, the ram pressure does not produce a force perpendicular to the wind to bend the tail more than 90 degrees like the Omega shape. If the Omega-tail galaxy is in orbital motion around the cluster center, the tails may appear to bend on its orbit. But in this case, the two tails should bend in the same direction along the orbit. Thus, both the north and south tails are bent inward to the core of the jet is not explained by the orbital effect. In this subsection, we discuss how the "Omega" structure was formed, considering three scenarios: the backflow of the jets, the Kelvin-Helmholtz instability (KHI), and the twin vortex. Whichever the case is, a re-acceleration mechanism is needed to make the Omega-shape structure visible in radio, as discussed in the previous section.


Numerical simulation \citep{2023ApJ...951...76O} indicates that a part of the jetted gas is reflected at the terminal shock, and strong backflow is made at an early stage of the jet ejection. Since this backflow is hotter and lighter than the original jet component, it can be trailed down due to the ICM wind. This may look like an Omega-shaped structure and could explain the origin of the observed one. The simulation also predicts that the backflow component is connected to the opposite jet component and forms a structure called a thread. Such a thread seems to be present in another head-tail galaxy in the Abell~3528 cluster \citep{2025A&A...694A..28D}, but has not yet been discovered in the observed head-tail galaxy. Future observation with high sensitivity will allow us to explore the thread, and we will conclude whether the backflow is the main cause of the Omega-shaped structure. 


When the jet is propagating into the ICM, shear motion is expected at the interface between the jet and the ambient ICM, developing KHI at the interface. The growth timescale of KHI can be scaled by the eddy turnover time, which is the timescale for one rotation of the vortex, ${\tau }_{\mathrm{eddy}}\,\sim {l}_{\rm t}/{v}_{k,\min }\sim {l}_{{\rm{b}}}/{v}_{{\rm{w}}}$, where ${l}_{\rm t}$ is the eddy size, ${v}_{k,\min}$ is the minimum mode $k$ (maximum size) of growing instability, $l_b$ is the jet length, and ${v}_{\rm w}$ is the wind velocity of the ambient ICM. The eddy turnover time of a jet with a length $l_b$ of about 200 kpc is ${\tau }_{\mathrm{eddy}} \sim 200$ Myr, if the velocity of the wind from the ICM is ${v}_{\rm w} \sim 1000$~km~s$^{-1}$. This timescale is consistent with the formation timescale of WAT, so that the jet bending can be related to the growth of KHI. 

Meanwhile, the KHI is more difficult to explain the formation of the tails and the Omega-shape structure for the following two reasons. Firstly, the eddy turnover timescale ${l}_{{\rm{t}}}$ of the terminal of the tail, i.e. lobe, which is about 30 arcsec or 100 kpc in diameter, is about half of ${l}_{{\rm{b}}}$ estimated above, suggesting that the KHI may not be well evolved. The southern tail is somewhat larger, but the conclusion does not change significantly. Secondly, even if the KHI structure is well developed, it tends to make an outward "TSUNAMI (GREAT WAVE)" component. Such a large-scale evolved component may explain the splash structure of the Omega-tail galaxy, but it appears at the opposite side of the bending direction of the Omega structure and does not contribute to the bending more than 90 degrees.


From the inward bending of the jets, one may recall a vortex. Because of the high density and high temperature of the cluster center, the Reynolds number increases rapidly toward the center. This means that, if the observed head-tail galaxy is moving toward the cluster center, the galaxy tends to destabilize its gas structure rapidly as it falls to the center. This implies that we are looking at the Omega-tail galaxy at the very moment of destabilization. The observed Omega structure may correspond to the vortex in this context, where the direction of the vortex is expected to be the same as the bending direction of the Omega structure.

A fine-tuning of the Reynolds number, however, will be required to produce the Omega-shape structure. It would be rather difficult to form a large-scale coherent component right behind the obstacle (galaxy) with a very large Reynolds number. With an intermediate Reynolds number, the Karman Vortices are regularly formed from both sides of jets, but they emerge alternately, and it is not consistent that both north and south tails bend over 90 degrees at the same time. Therefore, it is more likely that the observed Omega structure is related to the so-called ``twin vortex", which emerges around the critical Reynolds number. Even with this case, it is not clear whether the tail and splash structures can be solely produced by the vortex.

Although the critical Reynolds number of the ICM is not clear, let us adopt the Reynolds number $R_e = 46$ known as the critical Reynolds number for an incompressible fluid. Then, we can evaluate the kinematic viscosity $\mu$ from the definition, $R_e = \rho v^2/(\mu v/L)$. Given the typical values of galaxy clusters and the characteristic flow scale of $L=200$kpc, we obtain $\mu \sim 2.2 \times 10^3$~${\rm (g~ cm^{-1} s^{-1})}$, which is close to the Spitzer value. This number is also consistent with the previous observational work with Hitomi \citep{2021MNRAS.504.2800I}.

\section{Summary}\label{sec:5}

We conducted multi-frequency radio observations of Abell 3322 using the uGMRT and ATCA. The uGMRT observations were carried out in Band 3 (300-500 MHz). The analysis of the radio data yielded total intensity maps and spectral index maps of the Omega-tail galaxy. The total intensity maps reveal the core, jet, and tail structures of the Omega-tail galaxy with a high signal-to-noise ratio. Notably, we detected extended structures beyond the tail. The Omega-tail galaxy exhibits a unique "Omega" structure, where the tails bend more than 90 degrees from the jets. Furthermore, a lobe structure is observed at the tip of the tail, indicating an increase in intensity similar to that seen in AGN lobes.

The analysis reveals that the spectral index varies along the jet and tail, with a flatter spectrum in the core and steeper spectra downstream. This indicates energy losses as electrons propagate away from the core. Spectral modeling using KP, JP, and CI models reveals that the JP model provides the best fit in many regions. The spectral age is calculated, and the trend shows that the spectral age evolves more slowly than the sound speed propagation, suggesting re-acceleration or anti-aging effects. Comparison with simulations from \citet{2023ApJ...951...76O} shows that the observed re-acceleration efficiency is likely less than \FirstRevise{$\eta = 0.01$}. In conclusion, spectral aging analysis suggests that re-acceleration mechanisms are active. This study demonstrates the importance of spectral aging studies in understanding the interaction between AGN jets and the ICM.

We explore several possible explanations for this unusual morphology, including Kelvin-Helmholtz instability (KHI), Karman vortex formation, and backflow of the jets. KHI is considered as the jet moves through the ICM, creating velocity differences at the boundary, potentially leading to the formation of a WAT structure. However, the observed structure doesn't fully align with expected KH wave development, which should primarily create outward "splash" features. The Karman vortex is also explored, with the galaxy destabilizing as it falls toward the cluster center. While this could create an Omega-like structure, the simultaneous inward bending of both the north and south tails is inconsistent with the alternating nature of Karman vortex shedding. Furthermore, the coexistence of a tail and splash structure contradicts the strong sweeping action expected from a Karman vortex. The backflow of jets is then considered as a theory. As the early jet ejecting undergoes reflected backflow, it could generate the "collapsed" Omega structure. However, the presence of a tail component contradicts this scenario, and furthermore, there is no thread component found.

\begin{ack}
This work was supported in part by JSPS KAKENHI Grant Numbers 17H01110(TA), 21H01135(TA), and 25KJ0162(KK). This paper employs a list of Chandra datasets, obtained by the Chandra X-ray Observatory, contained in the Chandra Data Collection (CDC).
Data analysis was carried out on the Multi-wavelength Data Analysis System operated by the Astronomy Data Center (ADC), National Astronomical Observatory of Japan. 
\end{ack}

\bibliographystyle{apj}
\bibliography{export-bibtex}

\end{document}